\documentclass[aps,prl,twocolumn,amsmath,amssymb,superscriptaddress,showpacs%
]{revtex4}
\pdfoutput=1
\usepackage{amssymb,amsmath}
\usepackage{graphicx}
\usepackage{epsfig}

\newcommand{\be}{\begin{equation}}
\newcommand{\ee}{\end{equation}}
\newcommand{\req}[1]{Eq.~(\ref{#1})}
\newcommand{\reqs}[1]{Eqs.~(\ref{#1})}
\newcommand{\rref}[1]{(\ref{#1})}
\newcommand{\te}{t_{\text{el}}}
\newcommand{\tp}{t_{\text{ph}}}

\begin{document}

\title{{Jumps in  current-voltage characteristics in disordered films}}
 \author {Boris L. Altshuler}
  \affiliation{Physics Department, Columbia University,
538 West 120th Street, New York, N.Y. 10027, USA}
\affiliation{NEC-Laboratories America, Inc., 4
Independence Way, Princeton, N.J. 085540, USA}
  \author{Vladimir E.~Kravtsov}
   \affiliation{The Abdus Salam International Centre
for Theoretical Physics, P.O. Box 586, 34100 Trieste, Italy}
  \author{Igor  V.~Lerner}
 \affiliation{School of Physics and Astronomy, University of
Birmingham, Birmingham B15 2TT, United Kingdom}

\author{Igor L. Aleiner}
 \affiliation{Physics Department, Columbia University,
538 West 120th Street, New York, N.Y. 10027, USA}

\begin{abstract}
  We argue that giant jumps of  current at finite voltages observed
  in disordered films of {\em InO}, {\em TiN} and {\em YSi} manifest a bistability
  caused by the overheating of electrons. One of the stable states is
  overheated and thus low-resistive, while the other, high-resistive
  state is heated much less by the same voltage. The bistability
  occurs provided that cooling of electrons is inefficient and  the temperature dependence of the {\em equilibrium}
  resistance, $R(T)$ is steep enough. We use {experimental $R(T)$} and assume phonon mechanism of the
  cooling taking into account its strong suppression by disorder. Our
  description of  details of the $I-V$ characteristics does not involve
  adjustable parameters and turns out to be in a quantitative
  agreement with the experiments. We propose experiments for more
  direct checks of this physical picture.
\end{abstract}

\maketitle
\addtolength{\leftmargini}{-3mm}

Recent experiments on {{the}} $I$-$V$ characteristics of disordered films of $InO$
\cite{Shahar:05, Shahar2} and {\em TiN}  \cite{Bat1,Bat2}
(Similar findings were reported earlier
for amorphous {\em YSi} films \cite{Sanquer+}) are quite
intriguing.
Ohmic resistance $R$ of these films at temperatures $T < 1K $
demonstrates insulating behavior fit  by Arrhenius law ($k_B=1$)
\begin{align}\label{Arr}
    R(T)=R_0\exp\left[\left({\Delta}/{T}\right)^\gamma\right], \quad \gamma=1,
\end{align}
with $R_0$ and {$\Delta\simeq 1\div 5K$} being $T$-independent. Though this behavior is quite
interesting by itself \cite{KowOva:94}, it is the $I$-$V$ characteristics
 at low temperatures which turned out to be most unusual.
 When the voltage $V$
  is increased from $V=0$, the current
${\cal I}$ first increases
gradually, remaining rather small
  [highly resistive (HR) state].
 At a certain voltage
$V_{\text{HL}}$,
${\cal {I}}$
 jumps up by several orders of magnitude
  and  a low resistive (LR) state arises.
When the voltage is decreased
 from  $V > V_{\text{HL}}$, an inverse {jump} between the LR and HR states occurs
at a voltage $V_{\text{LH}} < V_{\text {HL}}$. These HR-LR switches persist
in a wide range of magnetic fields $B\simeq 0\div 10T$, with
 the {threshold} voltages $V_{\text{LH(HL)}}$  increasing
  with $B$.  {Authors of Refs.~\cite{Bat2,hahaha}
assumed that the HR state is a  new collective state  --
 ``superinsulator''.
Their calculation  was vigorously disputed in Ref.~\cite{FEW}.
}

In this Letter we show that these  {\em I-V} characteristics
can be explained without using  new concepts.
Our phenomenological approach predicts exactly this behavior when $T$-dependence
of the resistance  is steep as in Eq.~(\ref{Arr}) and the
electron-phonon ({e-ph}) thermalization is inefficient at low $T$.
We assume that {(similar assumptions for
  a system without
the switches were made in Ref.~\cite{Gershenson})}
\begin{enumerate}\setlength{\itemsep}{-1pt}
\item
 The electron-electron (e-e) interaction is
strong enough for electrons being mutually thermalized,
i.e.\ one can introduce their temperature   $T_{\text{el}}$
although the system is driven out of equilibrium by a finite voltage;
\item
 The e-ph interaction is weak, so that electrons can be out of
equilibrium with phonons (or any other thermal bath) of
temperature $T_{\text{ph}}$, i.e.\  $T_{\text{el}}> T_{\text{ph}}  $;
\item
 {$R(T)$-dependence   at a finite voltage is the same} as in
the ohmic regime, but $T_{\text{el}}(V)$ is substituted for $T$.
\end{enumerate}

These assumptions are sufficient to explain qualitatively the
experimental $I-V$ characteristics, {most carefully investigated in the accompanying Letter \cite{Shahar2}.}

{
{\em Phenomenological analysis} -- Temperature $T_{\text{el}}$
acquired by the electron due to a fixed external voltage $V$ is
determined by the  balance between the Joule heating  and the cooling
by the phonon bath (electron-electron collisions conserve the energy
and do not affect the heat balance):
Before specifying the model for the electron-phonon (e-ph) coupling we analyze the
cooling using model-independent arguments.
{The heat balance
equation for a sample of volume ${\mathcal
  V}$ and the electron density of states $\nu$ (assumed to be smooth at the Fermi level)
is
\begin{subequations}\label{1all}
 {\begin{align}\label{1}
&\frac{V^{2}}{R(T_{\text{el}})}=
\frac{{\cal E}(T_{\text {el}})}{\tau_{\text{e-ph}}(T_{\text {el}})}
-\frac{{\cal E}(T_{\text {ph}})}{\tau_{\text{e-ph}}(T_{\text {ph}})},
\\
&
\frac{1}
{\tau_{\text{e-ph}}(T)}
= \left(
\frac{T}{\hbar}
\right) \left(\frac{T}{\Theta}\right)^{\beta-3},
\label{1a}
\end{align}}
\end{subequations}
{where  ${\cal E}={\pi^2\nu {\mathcal V} T^2}/{6}$.
The  r.h.s of \req{1} obeys the following requirements:
(i) it vanishes in the thermal equilibrium; (ii) it provides the entropy
growth; (iii) at $T_{\text {el}}\gg T_{\text {ph}}$, it ceases to depend
on $T_{\text {ph}}$ as under such condition  the radiation of the
hot phonons by hot electrons
dominates.
Factor $T/\hbar$ provides
the proper dimensionality for the relaxation rate.}
The
second factor in r.h.s. of \req{1a}
describes a suppression of the e-ph coupling at
low $T$.
As the phonons are gapless, the suppression is
a power law.
 The energy
scale $\Theta$ encodes the strength of e-ph coupling.
We will specify  $\beta$ and $\Theta$
 later.
}

For a steep
 enough $R(T)$, there is a region on the $V$-$T_{\text{ph}}$ plane
 (for $T_{\text{ph}}<T_{\text{ph}}^{\text{cr}}$ and $V^<<V<V^>$) where
 Eqs.~(\ref{1all}) have two stable solutions: a ``hot" LH state with
 relatively high $T_{\text{el}}$ and a ``cold" HR state with
 $T_{\text{el}}\approx T_{\text{ph}}$.  Such a
 situation was discussed in Ref.~\cite{BAA2} in connection with the
 many-body localization \cite{BAA1}.

{
To analyze \reqs{Arr} -- \rref{1all} we use  dimensionless variables
\begin{align}\label{dim}
    t_{\text{el,\ ph}}
\equiv\frac{T_{\text{el,\ ph}}}{\Delta}\,,\quad v\equiv\frac{V}{V_0}\,,
\end{align}
and rewrite \req{1} in the form
 \begin{align}\label{ball1}
F(\te,\tp)=v;
\ F(x,y)\equiv\left[{\text{e}}^{1/x^{\gamma}}\left(x^\beta-y^\beta\right)\right]^{1/2}.
 \end{align}
The new characteristic voltage, $V_0$ emerging in \req{dim},
is natural to define through the electric
field scale $V_0 /L$, independent of the system size:
{
\be
\frac{V_0}{L}=\left(\frac{\pi^2 \nu \Delta^3}{6\sigma_0 \hbar}\right)^{1/2}
\left(\frac{\Delta}{\Theta}\right)^{\frac{\beta-3}{2}};
\quad \sigma_0=\frac{L^2}{R_0{\cal V}}.
\label{V0}
\ee}
The first factor in the
r.h.s. of \req{V0}
is uniquely determined as
a combination of local characteristics of an isolated system,
{$\nu,\ \sigma_0$} and $\Delta$, with a correct dimensionality.}
The problem of
nonlinear dissipative transport is meaningless without coupling of the
electrons with a thermal bath, {\em i.e.} when $\Theta \to
\infty$. This is reflected by the second factor in \req{V0}.

{
At the bistability boundaries $t_{\text{el}}^{\text{h,c}}$, the
derivatives
with respect to $t_\text{el} $ of both sides of  Eq.~(\ref{ball1}) are
equal, {\em i.e}:
\begin{align}\label{sol}\left(\beta/\gamma\right) t_{\text{el}}^\gamma=1- \left(
    t_{\text{ph}}/t_{\text{el}} \right)^\beta \,,
\end{align}
This equation has two positive solutions
if $\tp \leq
t_{\text{ph}}^{{\text{cr}}}(\beta)$. Here, the critical phonon
temperature,
$t_{\text{ph}}^{{\text{cr}}}=T_{\text{ph}}^{{\text{cr}}}/\Delta$
is
\be
t_{\text{ph}}^{{\text{cr}}}=\left(1+\beta/\gamma\right)^{
-\frac{\beta+\gamma}{\gamma\beta}}
< 1.
\label{tphc}
\ee
}

\begin{figure}[t]
\includegraphics[width=0.87\columnwidth,
]{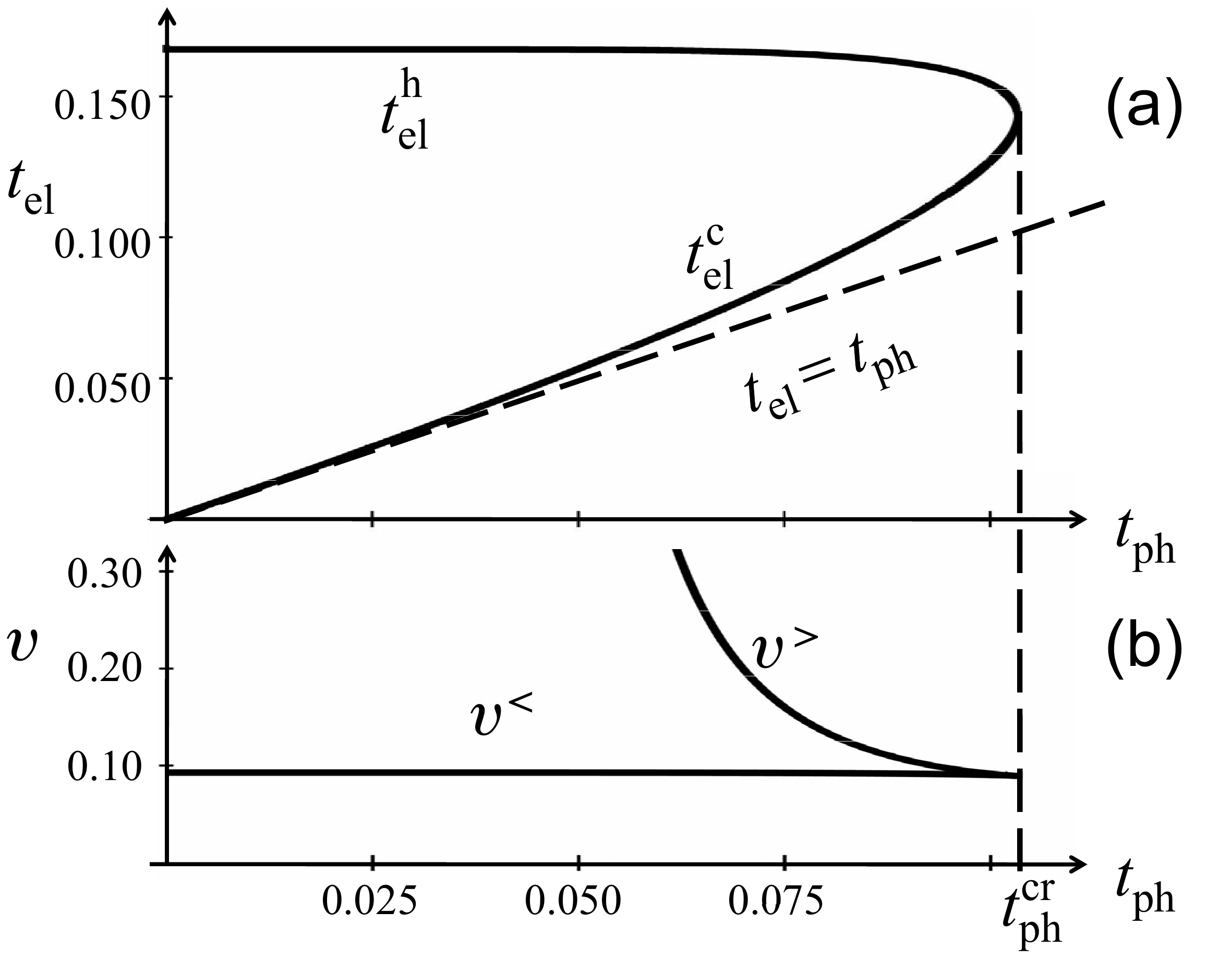}
\caption{Dependence of the dimensionless electronic temperature (a)
  and voltage (b) bistability boundaries on the bath temperature
  $t_{\text{ph}}$. Here $t_{\text{el}}^{\text{c}}$ and $v^>$ are the
  boundaries for the HR (``cold'') state and $t_{\text{el}}^{\text{h}}$
  and $v^<$ for the LR (``hot'') state.
{Both plots are for $\beta=6,\ \gamma=1$.}
}\label{fig2}
\end{figure}

{
The two solutions $\te^{\text{c(h)}}(\tp)$
of \req{sol}, see Fig.~\ref{fig2},  define the region of voltages $v^< <v < v^>$
where two states are locally stable, $
v^{<,>}=F\left(\te^{\text{h,c}},\tp\right)$.
\begin{subequations}
The  ``hot'' branch $ t_{\text{el}}^{\text{h}}$ turns out to only
slightly depend on $\tp$:
\be
\left(\frac{\gamma}{\beta+\gamma}\right)^{\frac{1}{\gamma}}\!\!
=\te^{\text{h}}(\tp^{\text{cr}})
\leq \te^{\text{h}}(\tp) \leq \te^{\text{h}}(0)
=\left(\frac{\gamma}{\beta}\right)^{\frac{1}{\gamma}}\!\!.
\ee
As the result,  $v^{<}(t_{\text{ph}})$ is almost
independent of $T_{ph}$,
\be
\label{v<}
\sqrt{\text{e}}
\left(\frac{\beta}{\beta+\gamma}\right)^\frac{\beta+\gamma}{2\gamma}
\left(\frac{\text{e}\gamma}{\beta}\right)^{\frac{\beta}{2\gamma}}
\leq v^< (\tp)
\leq \left(\frac{\text{e}\gamma}{\beta}\right)^{\frac{\beta}{2\gamma}}.
\ee
Contrarily, for $\beta \gg 1$, the ``cold'' branch
may be well approximated (except narrow vicinity of
$t_{\text{ph}}^{{\text{cr}}}$) by
\be\label{tcold}
\te^{\text{c}}- \tp \approx  \tp^{\gamma+1}/\gamma
< \tp/{\beta}.
\ee
For the upper bistability boundary, we find
\be
v^{>}(\tp)\approx \left[{\beta}/({e\gamma})\right]^{1/2}
\tp^{{(\beta+\gamma)}/{2}}
\exp\left[{1}/({2\tp^\gamma})\right].
\label{v>}
\ee
\label{vs}
\end{subequations}
}

{At each voltage in the bistability interval $v^< < v < v^>$ one of
the states is metastable.
As it is usual for the first order phase transition the voltages,
where the switches between HR and LR states happen (${V_{\text{HL}}}$ for HR$\to$LR
and  ${V_{\text{LH}}}$  for LR$\to$HR switches), are determined by kinetics of the
decay of metastable states. Theoretical analysis of this decay and
evaluation of ${V_{\text{HL,LH}}}$ is beyond the scope of this
paper. Here we
can predict only their bounds
\be\label{bounds}
{V_0}v^<=V^<  < {V_{\text{LH}}} < {V_{\text{HL}}} <V^>={V_0} v^>.
\ee
The difference between the local instability and
the metastable state decay transition can be
ascertained from  the slope $d{\cal I}/dV$ near the
transition,
see Fig.~\ref{fig:new}.
}

Finally, it is important to emphasize that \reqs{dim} -- \rref{V0}
imply a non-trivial scaling of the bistability bounds,
\be
\label{scaling}
V^{<,>}=\Delta^{\beta/2}f^{<,>}\left(T_{\text{ph}}/\Delta\right),
\ee
which should apply to ${V_{\text{HL, LH}}}$
provided that the switches are close to (a)-type, see Fig.~\ref{fig:new}.

\begin{figure}[b]

\includegraphics[width=1\columnwidth,
]{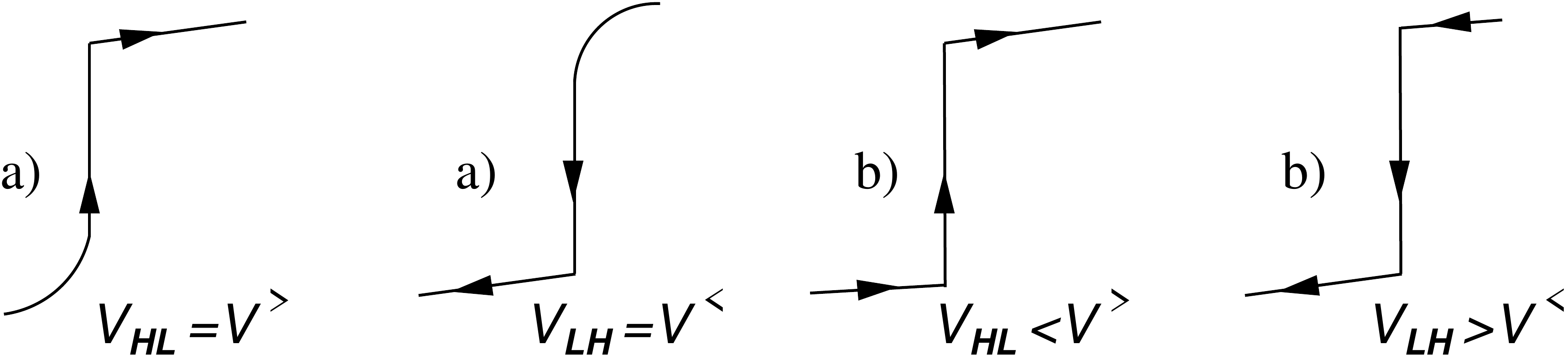}
\caption{The schematic $I-V$ characteristics for the
switches occuring through the local instability [type (a)] and
through the decay of metastable state [type (b)].}
\label{fig:new}
\end{figure}

{
The $I-V$ characteristics is most conveniently visualized by
calculating the non-linear conductance $G$
\be\label{nlG}
{{\cal I}=G(V,T_{\text{ph}})}{V}; \quad
G(T_{\text{ph}},V)=\frac{\exp\left[{-\te(v)^{-\gamma}}\right]}{R_0},
\ee
see \req{Arr}, where $t_{\text{el}}(v)$ is found from \req{ball1}.
The $I-V$ curves obtained by substituting the numerical solution of \req{ball1} into
\req{nlG} are plotted on Fig.~\ref{fig4}. The HR and LR
branches are connected by  unstable branches shown
by dotted lines. The HR$\to$LR {(``cold'' to ``hot'' state)} and
LR$\to$HR {(``hot'' to ``cold'' state)} switches are depicted by vertical
dashed lines.

If $T_{\text{ph}}$ is not too close to the critical temperature
\rref{tphc},
we can use the fact $\te-\tp \ll \tp$, see \req{tcold}, to obtain
an analytic description of $G(V,T_{\text{ph}})$ for the HR state:
\be
\label{univ}
\left(\frac{v}{v^>}\right)^2=\frac{\text{e} \ln \left[GR(T_{\text{ph}})\right]}
{GR(T_{\text{ph}})}; \quad {1} \leq G {R(T_{\text{ph}})}
\leq {\text{e}}.
\ee
}
Note, that the dependence  is universal, i.e. it holds for arbitrary
 values of exponents $\beta,\gamma$.

{
{\em Microscopic input} --
To quantify the developed phenomenology we rely on
conventional theory of normal disordered metals
neither seeking  a microscopic explanation for, e.g., the
Arrhenius law \rref{Arr}
nor involving physics of the insulator-superconducting transition.
The current jumps occur in the insulating regime $T < \Delta$.
However, the thermal balance equation is
valid also for $T \gtrsim \Delta$ and
the e-ph coupling should be a continuous function.
Thus, our strategy is to use the theory of the e-ph interaction
in dirty metals \cite{Schmid,Reizer} for    $T > \Delta$ and
extrapolate to the lower temperatures. The cooling rate is determined
by the material mass density, $\rho$, and the transverse sound
velocity  $c_{\text{s}}$. Disorder is known to suppress the
e-ph coupling if the wave length of a thermal phonon exceeds the electron
elastic mean free path, ${\hbar c_{\text{s}}}/ T \gg \ell$, (as it
does for films of Refs.~\cite{Shahar:05,Shahar2,Bat1,Bat2,Sanquer+} where $c_{\text{s}}\sim
  3\times10^{5}\,{\rm cm/s}$,  $T\lesssim 1\,K$
and
$\ell\lesssim 10$nm).
The result is  \cite{Reizer}
\be \label{Etau}
\frac{{\cal E}(T)}{\tau_{\text{e-ph}}(T)}
=\frac{\alpha^2\,k_\text{F}\ell\,
   \mathcal{V}n_{\text{el}}\,T^6}{\hbar^4 \rho c_\text{s}^5 } \,,\quad
\alpha=\frac{\,2\pi^2}{\sqrt{315}}
   \approx1.1\,.
\ee
where $n_{\text{el}}$ is the conduction electrons density,
and $k_{\text{F}}=(3\pi^2n_{\text{el}})^{1/3}$ is the Fermi momentum \cite{footnote,footnote2}.
Comparing \reqs{Etau} and \rref{1a} we obtain
\be\label{betatheta}
\beta=6,\quad \Theta=\left(
\frac{\pi^2}{6\alpha^2}\frac{\hbar^3 \nu\rho c_\text{s}^5 }{n_{\text{el}}k_{\text{F}}\ell}
\right)^{1/3}.
\ee
Substituting \req{betatheta} into \req{V0}, and using, at $T > \Delta$, the Drude formula
$\sigma_0 = (e^2/\hbar)
(k_{\text{F}}{\ell})(n_{\text{el}}/k_{\text{F}}^2)$, we find
\be\label{V0micro}
\frac{eV_0}{L}=\alpha \frac{ k_{\text{F}}\Delta^3}{\left(\rho c_\text{s}^5\hbar ^3\right)^{1/2}}.
\ee
}{
Remarkably, the elastic mean free-path $\ell$ -- the only quantity,
which becomes meaningless in the insulator, does not enter electric
field scale $V_0/L$. It makes us believe that \req{V0micro} derived for the
metallic regime $T>\Delta$ may be extended to the insulator at $T<\Delta$.  }

\begin{figure}[t]
\includegraphics[width=.7\columnwidth,
]{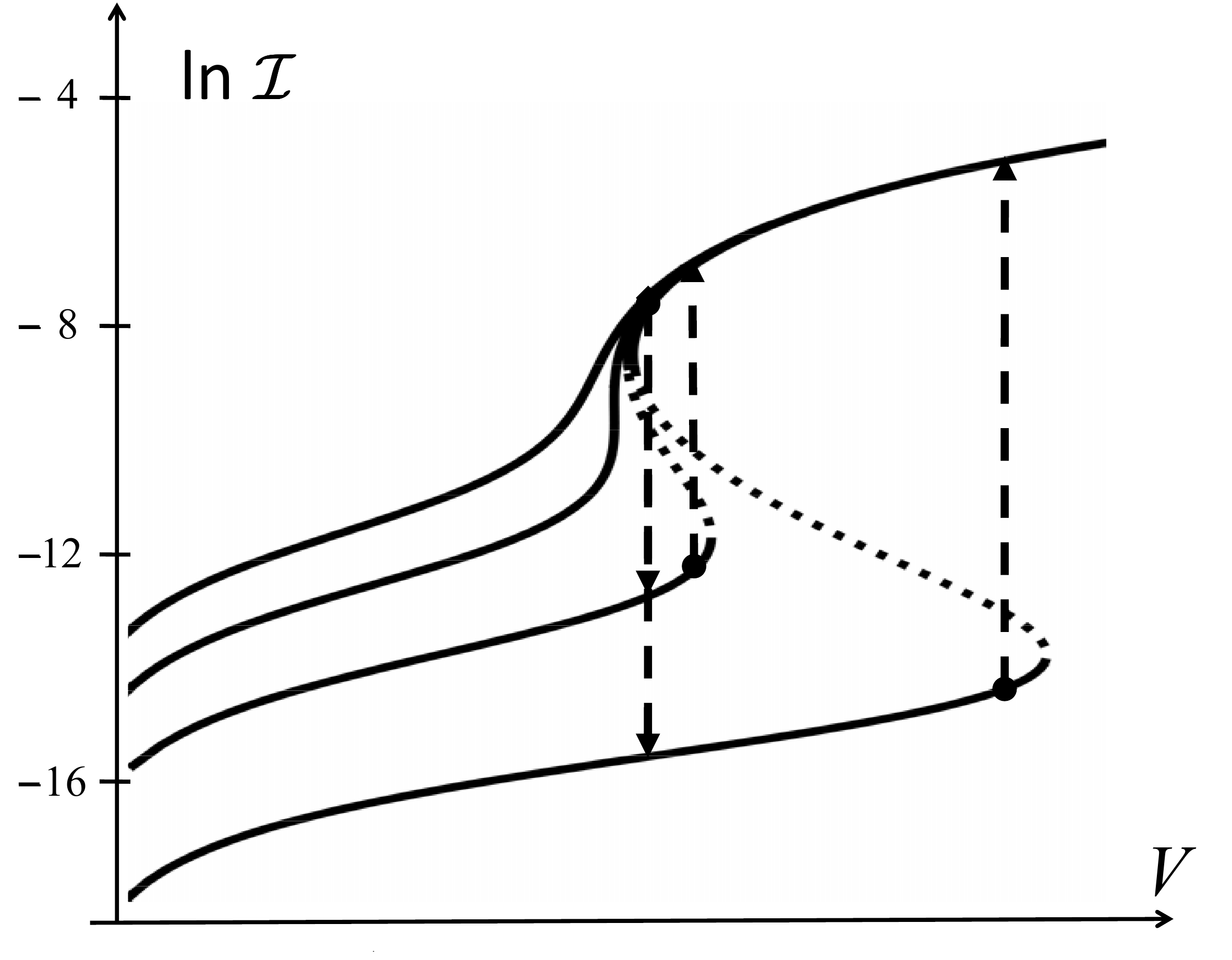} \caption{Universal I-V characteristics  for
$\beta=6,\ \gamma=1$
(going from the top) $t_\text{ph}= 1.15\,t_\text{ph}^\text{cr}$, and
$t_\text{ph}=t_\text{ph}^\text{cr}$, $t_\text{ph}=0.90 \,t_\text{ph}^\text{cr}$ and
$t_\text{ph}=0.75\,t_\text{ph}^\text{cr}$. Instable regions on the S-shaped curves are shown
by dotted lines which join the upper and
lower solid lines {at the stability boundaries} of the LR (``hot'') and HR (``cold'') state,
respectively. The vertical dashed lines with arrows show {possible positions for LR$\rightleftarrows$HR current jumps}. The units of $\mathcal{I}$ and $V$ are arbitrary, but the values of the current switches in the log-scale are universal and in a good agreement with the experimental values.   } \label{fig4}
\end{figure}

{\em Qualitative comparison with experiment}--
Several conclusions  of our phenomenological consideration
can be compared directly
with the experimental  results:
\begin{enumerate}
\setlength{\itemsep}{0pt}
\item
The $I-V$ curves in
\cite{Shahar:05,Shahar2,Bat1,Bat2,Sanquer+} look like those on Fig.\ref{fig4}
for both $T>T^{\text{cr}}_{\text{ph}}$ and
$T<T^{\text{cr}}_{\text{ph}}$.
At $T\to T^{\text{cr}}_{\text{ph}}+0$  the inflection point becomes increasingly
pronounced. At $T$ below $T^{\text{cr}}_{\text{ph}}$,
hysteretic HR$\to$ LR and LR$\to$ HR
current jumps occur. These jumps indeed reach several orders of magnitude;
\item
At a rather large
interval of voltages the dependence of $\log {\mathcal I}$ on $V$
looks like linear;
\item
$V_{LH}$ only slightly changes with $T_{\text{ph}}$, while
$V_{\text{HL}}$ substantially
increases when phonon temperature is reduced.
This is what  \reqs{v<}, \rref{v>} predict.
\item  Observed $T_{\text{ph}}$-dependence of
$V_{\text{HL}}$ is still weaker than the exponential dependence \rref{v>}. This is
consistent, however, with bound \rref{bounds} and the
type (b) HR$\leftrightarrow$LR transition (see discussion before
\req{bounds}, and Fig.~\ref{fig:new}).
{Another possible reason for a relatively slow dependence of $V_{HL}$ on $T_{ph}$ is a deviation from the Arrhenius law of Eq.~\rref{Arr}.}

\item
The assumption that $V_{\text{HL}}<V^>$ and $V_{\text{LH}}
 \approx
V^<$  implies  the
shapes of the $I-V$ characteristics to be of type (a) close to LR$\to$ HR switch
and of type (b) in the vicinity of HR$\to$ LR switch (see Fig.\ref{fig:new}).
This is exactly what was observed.
\end{enumerate}

{
Further analysis of data of Ref.~\cite{Shahar2} uncovers an
intriguing   discrepancy: the observed ratio of the nonlinear
conductances $G(V_{\text{HL}})/G(V\to 0)$, see \req{nlG},
is noticeably bigger than the bounds of \req{univ} and
 Fig.~\ref{fig4}, valid for any
single scale dependence $R(T)$.
This discrepancy can be resolved within our
approach only by involving an extra temperature scale $T^* < T^{\text{cr}}_
{\text{ph}}$ in addition to $\Delta$. While the  origin of this scale is
yet to be understood, the many-body localization \cite{BAA1} is a
possibility.
}

{{\em Quantitative discussion}.
The  magnitude of the LR-HR jump is almost independent of other experimental
    parameters.  Since locations of the jumps between the upper
    and lower stability boundaries are ill-defined, we can only
    estimate  the order of magnitude. For $T_{\text{ph}}=0.75
  T^{\text{cr}}_{\text{ph}}$ {the predicted $4$-$5$ orders in
    magnitude of the current jump} (Fig.~\ref{fig4}) agrees reasonably
  with experiment.   {Experimental cooling is well fitted with
    $\beta=6$ in Eq.~(\ref{1all}),
in agreement with the above description of the electron-phonon
mechanism}. The critical bath
temperature can be estimated from the temperature scale
{$\Delta\approx1.2$K: although the Arrhenius
law  does not give a very good fit,  this scale can be rather reliably
extracted from the data}\cite{Shahar2}.
Then for $\gamma=1$ and $\beta=6$, we find
 $
    T_{\text{ph}}^{{\text{cr}}}\approx 0.1 \Delta\approx 0.1\textrm{K}
$   in agreement with the experimental values
\cite{Shahar2}.}

{More quanitative comparison is hindered by the strong sensitivity of
the critical temeperature and switching voltages to the choice of
$\gamma$, see \reqs{tphc} -- \req{vs}.
For a consistent \textit{quantitative} comparison with
  experimental observation,
one should fit the experimentally observed equilibrium resistance $R(T)$ into the left-hand-side of the heat balance equation Eq.~(\ref{1}) and solve the resulting equation numerically. This is done in the accompanying experimental Letter \cite{Shahar2}.}

{{\em In conclusion,} there is a number of strong evidences in favor of the
  electron overheating being the main cause of the giant HR$\leftrightarrow$LR current
  jumps observed in \cite{{Shahar:05,Shahar2,Bat1,Bat2,Sanquer+}}.  Direct
  detection of the electron overheating through e.g., noise
  measurements would be an unambiguous proof of this.} As the cooling
is a rather slow process it looks plausible to perform the time
resolved-studies of  electron transport, {\em e.g.} measure the
current caused  by a train of voltage pulses as a function of the
pulse duty cycle. Such a measurement could also shed some light
on the kinetics of the switches. Also, the scaling relation
\rref{scaling}
is very characteristic for the overheating mechanism
(it can be verified by tuning $\Delta$, e.g., by
gentle annealing).

Finally, we emphasize the importance of our interpretation of  the
data \cite{{Shahar:05,Shahar2,Bat1,Bat2,Sanquer+}}
in the context of the general theory of the electron transport.
Overheating of the electrons is quite usual in low
resistive metals
{\cite{VolKog}}. {As to insulators (resistance far in excess of
$h/e^2$) the overheating was rarely \cite{Gershenson} considered
quantitatively, because the
conventional mechanisms of the low temperature charge transport are
based on phonon-assisted hopping (see, e.g., \cite{hopping}).} We fully realize that our
explanation of the current jumps contradicts  the standard
picture (as well as the Arrhenius law \req{Arr}, $\gamma =1$, though). At
the same time, we find the arguments in favor for the overheating being
observed in Refs.~\cite{Shahar:05,Shahar2,Bat1,Bat2,Sanquer+} to be
quite convincing. If our explanation is confirmed, these
experiments should be considered as the first (to the best of our
knowledge) reliable evidence of the {strong} overheating in the insulating
state and, thus, of
the existence of phonon-independent transport in insulators.

\begin{acknowledgments}
We thank M.~Ovadia, B.~Sacepe and D.~Shahar for sharing with us their
experimental results \cite{Shahar2} prior to publication and
Yu.~M.~Galperin for reading the manuscript and valuable remarks.
  V.~E.~K.\ is grateful
to M.~V.~Feigel'man for stimulating discussions.  {V.~E.~K.\ and
  I.~V.~L.\ acknowledge kind hospitality
 extended to them at the Newton Institute (Cambridge).}  We acknowledge support by
EPSRC grant T23725/01, and by the US DOE contract No.\ DE-AC02-06CH11357.
\end{acknowledgments}


\begin{references}
\bibitem{footnote2}{Equation \rref{Etau} is obtained within a rather crude
  jelly model. Even though this does not affect the value of the exponent
  $\beta=6$, the value of $\Theta$ in \req{betatheta} may be off by
  an unknown numerical factor.}
\end{references}
\end{document}